\documentclass[prb,twocolumn]{revtex4-1}

\usepackage{graphicx}
\usepackage{dcolumn}
\usepackage{bm}
\usepackage{amsmath}

\begin{document}
\preprint{}

\title{Observation of Magnonic Band Gaps in Magnonic Crystals with Nopnreciprocal Dispersion Relation}
\author{M. Mruczkiewicz$^{1}$, E. S. Pavlov$^{2}$, S. L. Vysotsky$^{2}$, M. Krawczyk$^{1}$, Yu. A. Filimonov$^{2,3}$, S. A. Nikitov$^{3,4,5}$}
\address{$^{1}$ Faculty of Physics, Adam Mickiewicz University in Poznan, Umultowska 85, Pozna\'{n}, 61-614 Poland\\
$^{2}$ Kotel'nikov Institute of Radio-Engineering and Electronics of RAS, Saratov Branch, Zelenaya Str. 38, Saratov, 410019, Russia. \\
$^{3}$ Saratov State University, Astrakhanskaya Str. 83, Saratov, 410012, Russia. \\
$^{4}$ Kotel'nikov Institute of Radio-Engineering and Electronics of RAS, Mokhovaya Str. 11, Bld. 7, Moscow, 125009, Russia.\\
$^{5}$ Moscow Institute of Physics and Technology, Dolgoprudny, 141700, Moscow Region, Russia.
}

\date{\today}
\begin{abstract}

An effect of metallization of the magnonic crystal surface on the band gaps formation in the spectra of the surface spin wave (SSW) is studied both theoretically and experimentally. The structures under consideration are one-dimensional magnonic crystals based on yttrium iron garnet with an array of etched grooves with metal screen on the top of the corrugated surface and without it. Due to nonreciprocity of propagation of the SSW the shift of band gap to higher frequency and from the border of the Brillouin zone in presence of conducting overlayer was measured in transmission line experiment. Results of  numerical calculations and model analysis are in agreement with experimental data and give further insight into origin of the band gap and properties of the nonreciprocal SSW in metallized magnonic crystals. This gives positive answer to the outstanding question about possibility of detection of magnonic band gaps in the spectra of the spin waves with nonreciprocal dispersion in magnonic crystals and creates potential for new applications and improvements of already existing prototype magnonic devices.


\end{abstract}

\pacs{75.30.Ds, 75.75.-c, 76.50.+g}
\maketitle

The surface spin waves (SSW) propagating in a ferromagnetic film along the perpendicular direction to a tangentially applied magnetic field \textbf{\textit{H}} (Damon-Eshbach geometry)\cite{DE} posses a nonreciprocal properties. More specifically, SSW amplitude distribution through the film thickness $d$ has maximum near opposite surfaces for waves with converse directions of the wavevector \textbf{\textit{k}} or bias field \textbf{\textit{H}}, while frequency $f$ of the oppositely directed waves are the same $f(k)=f(-k)$. Nonreciprocity in dispersion \cite{Seshadri70} $f(k) \neq f(-k)$ and attenuation $\text{Im} f(k) \neq \text{Im} f(-k)$ \cite{Wolfram70} of the oppositely directed SSW  can appear if metallic layer placed near one of the ferromagnetic film surfaces. The changes in dispersion of the SSW propagating along the metallized surface can be treated in approximation of  an influence of ideal metal if  thickness $D$ of the metallic layer satisfied the condition $D>3k \rho$ ,\cite{New4} where $\rho$ is the skin depth in metal at the SSW frequencies. In this case nonreciprocity in dispersion $f(k) \neq f(-k)$ manifests itself not only in changes of the dispersion slope of SSW propagating along metallized surface, but for sufficiently small thickness $t$ of the dielectric spacer between  metallic and magnetic layers ($t<2d$) leads to  expanding of the SSW bandwidth on some value $\Delta \Omega$ above the frequency range of SSW traveling along free surface. \cite{New5} At frequencies $\Delta \Omega$ propagation of SSW is unidirectional (possible only along metallized surface) that can be used for waves steering and channeling control \cite{New6} and suppression of the spatial resonances. \cite{New7} The SSW propagation in metallized magnetic films was also studied from the point of view the signal delay line design \cite{New8} and soliton formation. \cite{New9}


Since magnonic crystals (MCs) bring wide attention of researchers due to its  usefulness for manipulation of spin wave band structure,\cite{Krawczyk98,Neus08,Tachi12_PRB86,Krawczyk14,Barman14} it is natural that the  influence of metallization on MCs was also recently investigated. It was shown experimentally that the band gap at frequency of unmetallized structure disappears when the metal is attached.\cite{Beginin:252412} However, as yet the predictions of numerical calculations that the gaps appear at higher frequency range were not demonstrated experimentally.\cite{Sokolovskyy:07C515,Mruczkiewicz2013b} Here we present an experimental demonstration of the nonreciprocal dispersion relation of spin waves and existence of an indirect magnonic band gap in one dimensional magnonic crystals in contact with the metallic film.  

The nonreciprocal properties of magnonic band structure are potentially useful to design miniaturized microwave isolators and circulators, essential elements in microwave technology. The structure based on the uniform yttrium iron garnet (YIG) film with metallic stripes was proposed as a ultrasensitive magnetic field sensors.\cite{Inoue11:132511} Although nonconsidered in that study, the nonreciprocal properties of SSW can influence positively their sensitivity. The nonreciprocity can be also of big importance for magnonics, where spin waves are used to process information.\cite{Kruglyak10b,Khitun10} The dispersion of metal-YIG structures is also an important factor for electrically tunable magnonic device. \cite{Tang2014}  The  technologically simple design proposed here can be of advance over other complicated magnonic crystal based nonreciprocal structures.\cite{Verba:082407} Moreover, the influence of metal in the surrounded space of a magnonic device, especially when integrated, can aditionally result in changes of its functionality.\cite{Chumak10, patent3}

\begin{center}
\begin{figure}[!ht]
\includegraphics[width=8cm]{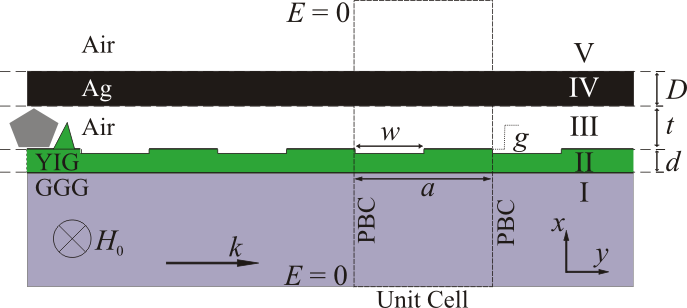}
\caption{(Color online) A schema of the structure under consideration. The materials are indicated as follows: I) dielectric substrate (GGG), II) magnonic crystal (YIG), III) air gap, IV) metallic overlayer (Cu) and V) air surrounding. The bias magnetic field $H_0$ is in the film plane and directed along the $z$ axis. The SSWs propagate along the $y$ axis. The rectangular unit cell used in numerical calculations is marked by dashed line. The periodic boundary conditions (PBC) were used along the $x$ axis. The bottom and top border of the unit cell is far from the structure, at these borders we assume fields equal 0.}
\label{structure}
\end{figure}
\end{center}

Here, we investigate an influence of a metal on the magnonic band structure in two microscale MCs made of YIG films with etched grooves. The structure under consideration is depicted in Fig. \ref{structure}.  The YIG films were grown on the $<$111$>$-oriented gadolinium gallium garnet (GGG) substrate. The grooves were introduced by chemical etching on the YIG surface, which form a periodic structurization. The first MC (Sample I) consists of the $d \approx 7.7$~$\mu$m thick YIG film with an array of etched grooves of the depth $g \approx 1.5$~$\mu$m and width $w \approx 80$~$\mu$m. The separation between grooves is approximately $70$~$\mu$m and the lattice constant is of $a \approx 150$~$\mu$m. The second MC (Sample II) is the $d \approx 3.8$~$\mu$m thick YIG film with an array of grooves with the depth: $g \approx 2$~$\mu$m and the width: $w \approx 50$~$\mu$m. The separation between grooves is approximately $150$~$\mu$m and the lattice constant is of $a \approx 200$~$\mu$m. The wavevectors at the edge of the first Brillouin zone (BZ) for the investigated structures are $2 \times 10^4$ 1/m and  $1.5 \times 10^4$ 1/m, for the sample I and II, respectively. On the basis of analysis conducted in Ref. ~[\onlinecite{New5,Mruczkiewicz2013b}]  we can expect that effect of metallization on the magnonic gap formation ought to be observable if thickness of the dielectric spacer $t$ is smaller than lattice constant $a$ ($t<a$). For chosen structures this requirement will be fulfilled if $t\approx 100$~$\mu$m.  On the other hand the air gap of such values can be easily realized experimentally. \cite{vysotskii2011effect,vysotskii2013bragg}

\begin{center} 
\begin{figure}[!ht]
\includegraphics[width=8cm]{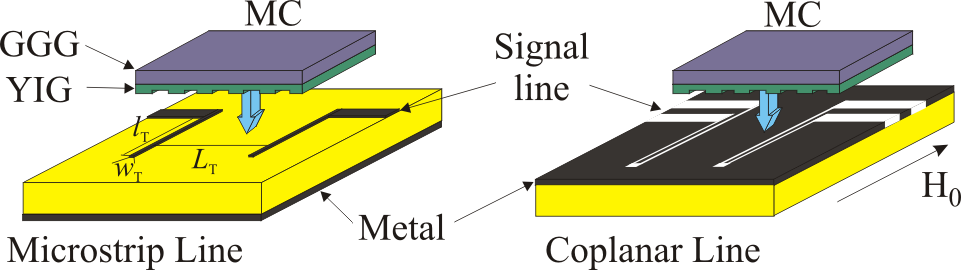}
\caption{(Color online) A microstrip (left) and coplanar (right) line used in the SSW transmission measurements through MC without and in contact with metallic overlayer. In the experimental setup MC sample is placed atop with groves side close to the transducers.}
\label{experiment}
\end{figure}
\end{center}

To investigate features of magnonic band formation in spectra of nonreciprocal waves
the measurements of transmission and phase characteristics in two types of SSW delay line based on MCs were performed. The first type was made by putting MCs on the microstripe transducers, while in the second type coplanar transducers were used, see Fig. \ref{experiment}.  Both types of transducers were prepared by using photolithography on alumina substrates covered by Cu layers with thickness $D \approx 5$ $\mu$m.  Input and output transducers have width $w_{\text{T}} \approx 30$ $\mu$m length $l_{\text{T}}\approx 4$~mm and were separated by distance $L_{\text{T}} \approx$ 4 mm. The YIG films under investigations had planar dimensions 15 mm~$\times$~5 mm. The whole structure was placed in the gap of an electromagnet so that the external magnetic field was oriented along the grooves and parallel to the transducers. The field was strong enough to saturate the sample ($\mu_0 H_0 \approx 41.6$ mT). The amplitude-frequency (AF) and phase-frequency (PF) characteristics of the delay line prototype models were measured by an Agilent E5071C-480 network analyzer. The ground layer of metal of the coplanar line serves also as a metal overlayer depicted in Fig. \ref{structure}. 

\begin{center}
\begin{figure}[!ht]
\includegraphics[width=0.45\textwidth]{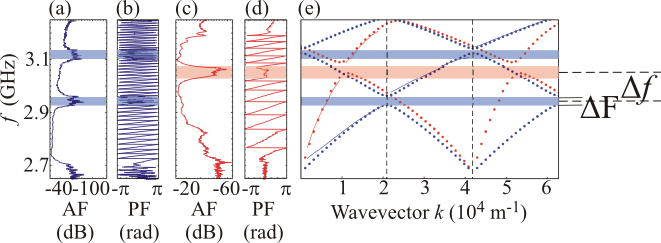}
\caption{(Color online) Result of the measurements for Sample I ($a=150$ $\mu$m). a) Amplitude-Frequency (AF), b) Phase-Frequency (PF) characteristics of YIG MC without metal overlayer, c) AF and d) PF characteristics of YIG MC with metal overlayer. e) The comparison of dispersion relation obtained from numerical calculations (blue dots for unmetallized sample, red dots for metallized) and  from measured PF signal (blue lines for unmetallized sample and red lines for metallized). The transparent blue and red bars indicate experimentally measured positions of magnonic band gaps in unmetallized and metallized sample, respectively. The vertical dashed lines in (e) indicate the edges of Brilouine zones, $\Delta f$ is the frequncy shift of the magnonic gap induced by metallization and $\Delta F$ is the magnonic bad gap width.}
\label{results1}
\end{figure}
\end{center}

\begin{center}
\begin{figure}[!ht]
\includegraphics[width=0.49\textwidth]{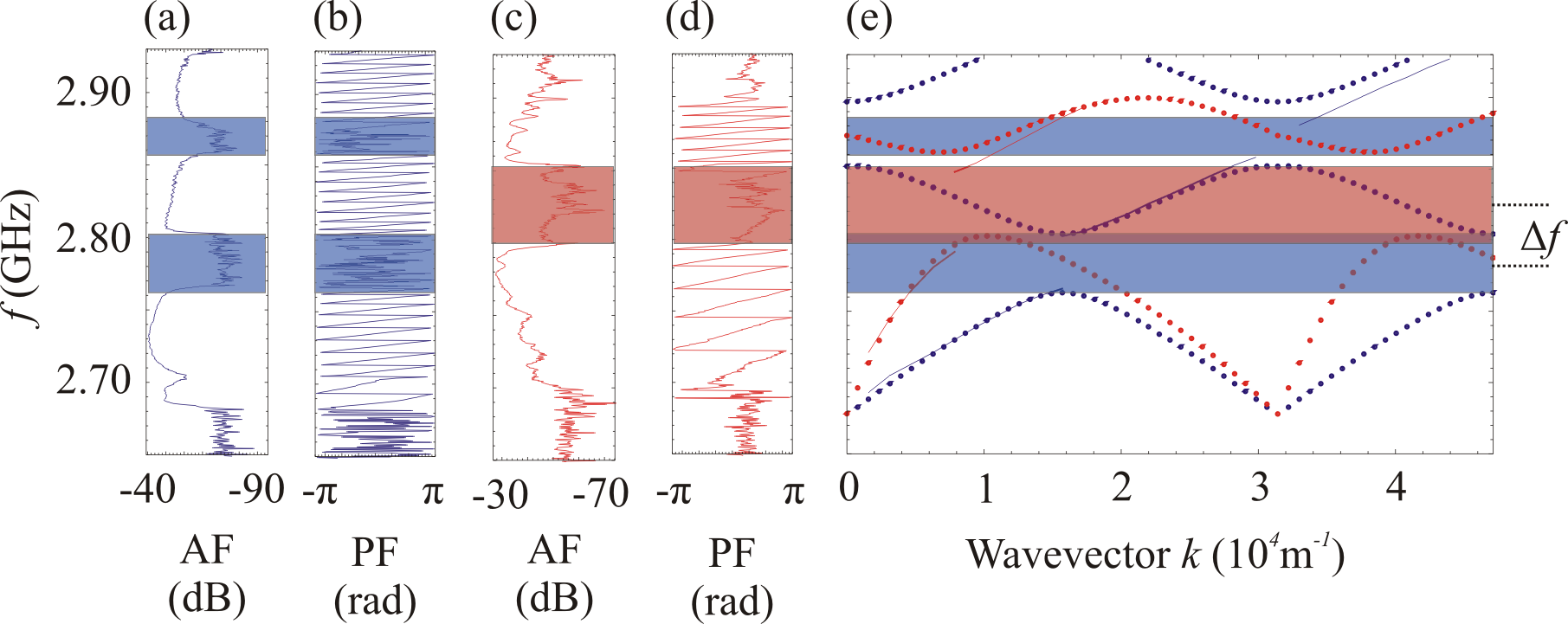}
\caption{(Color online) The result of measurements and calculations for Sample II ($a=200$ $\mu$m). (a) AF and (b) PF characteristics of YIG MC without metal overlayer; (c) AF and (d) PF characteristics of YIG MC with metal overlayer. (e) The comparison of dispersion relation obtained from numerical calculations (blue dots for unmetallized sample, red dots for metallized) and  from  PF signal (blue line for unmetallized sample, red line for metallized). The transparent blue and red bars indicate experimental positions of the magnonic band gaps in unmetallized and metallized sample, respectively.}
\label{results2}
\end{figure}
\end{center}

Using those two different types of delay lines (microstripe and coplanar) the transmission of SSWs in MC without and with a metallic overlayer were measured for Sample I and II. The measured transmission spectra (AF characteristic) are shown in Figs.~\ref{results1}(a) and (c) and in Figs.~\ref{results2}(a) and (c) for Sample I and II, respectively. Magnonic band gaps, $\Delta F$ indicated by low AF signal are found in both transmission spectra and marked by shadowed rectangles.  However, in the case of MCs with metallic overlayer  (Figs.~\ref{results1}(c) and \ref{results2}(c)) the first magnonic gap is shifted to the higher frequencies on some values $\Delta f$, see Fig. \ref{results1}(e) and Fig. \ref{results2}(e). The frequency shift was $\Delta f\approx120$ MHz and $\Delta f\approx140$ MHz for samples I and II, respectively. The existence of these gaps is confirmed also in the respective PF characteristics which are presented in Figs.~\ref{results1}(b) and (d) and Figs.~\ref{results2}(b) and (d) for Sample I and II, respectively.


In order to obtain more insight into the formation of the magnonic band gaps in the nonreciprocal structure the numerical calculations of the dispersion relation were performed. For SSWs from the GHz frequency range, a large value of thickness of MCs considered in the paper and a small value of exchange constant of YIG the  magnetostatic approximation is well justified. Therefore, the exchange coefficient in YIG is neglected in our theoretical investigation. In order to calculate the SSW dispersion relation we solved the wave equation for an electric field vector ${\bf E}$:\cite{raju2006antennas}
\begin{equation}
\nabla \times \left(\frac{1}{\hat{\mu}_{r}({\bf r})} \nabla \times {\bf E} \right) - \omega^2 \sqrt{\epsilon_{0}\mu_{0}} \left( \epsilon_{0}-\frac{i \sigma}{\omega	 \epsilon_{0}} \right) {\bf E}=0, \label{eq:main}
\end{equation}
where $\omega = 2 \pi f$, $\mu_{0}$ and $\epsilon_{0}$ denote the vacuum permeability and permittivity, respectively, and $\sigma$ is the conductivity, different from zero only in the metallic overlayer. To describe the dynamics of the microwave magnetization components perpendicular to the external magnnetic field, it is sufficient to solve the Eq. \ref{eq:main} for the $z$ component of the electric field vector ${\bf E}$ which depends solely on $x$ and $y$ coordinates \cite{mru2014}: $E_{z}(x,y)$. The MCs under consideration are assumed infinte along the $z$ direction. The permeability tensor~$\hat{\mu}({\bf r})$ is obtained from the linearized damping-free Landau-Lifshitz equation:\cite{Gur96}
\begin{equation}
\hat{\mu}_{r}({\bf r})=\left(
\begin{array}{ccc}
\mu^{xx}({\bf r}) & i\mu^{xy}({\bf r}) & 0\\
-i\mu^{yx}({\bf r}) & \mu^{yy}({\bf r}) & 0\\
0 & 0 & 1
\end{array}\right),
\end{equation}
where
\begin{eqnarray}
\mu^{xx}({\bf r}) &=& \frac{\gamma \mu_{0} H_{0}(\gamma \mu_{0} H_{0}+ \gamma \mu_{0}  M_{S}({\bf r}))-(2 \pi f)^2}{(\gamma \mu_{0} H_{0})^2-(2 \pi f)^2}, \\
\mu^{xy}({\bf r}) &=&  \frac{\gamma \mu_{0}  M_{S}({\bf r}) 2 \pi f}{(\gamma \mu_{0}H_{0} )^2-(2 \pi f)^2}, \\
\mu^{yx}({\bf r}) &=& -\mu^{xy}({\bf r}), \;\;\;\; \mu^{yy}({\bf r}) = \mu^{xx}({\bf r});
\end{eqnarray}
$M_{S}$ is the saturation magnetization of YIG and $\gamma$ is the gyromagnetic ratio. Eq.~(\ref{eq:main}) has coefficients being constants  or periodic function of $y$ (with a period $a$), thus the Bloch theorem can be used:
\begin{equation}
E_{z}(x,y)=E_{z}^\prime(x,y)\text{e}^{i k_{y} \cdot y },
\label{bloch}
\end{equation}
where $E_{z}^\prime(x,y)$ is a periodic function of $y$: $E_{z}^\prime(x,y) = E_{z}^\prime(x,y+a)$. $k_y$ is a wave vector component along $y$ and due to considering SSW propagation along this direction only we assume $k_y \equiv k$. Eq.~(\ref{eq:main}) together with  Eq.~(\ref{bloch}) can be written in the weak form \cite{weak} and the eigenvalue problem can be generated. This eigenequation is supplemented with the Dirichlet boundary conditions at the borders of the computational area placed far from ferromagnetic film along $x$ axis (dashed lines in Fig.~\ref{structure}). This eigenequation is solved by the finite element method with the use of a commercial COMSOL software.

In calculations we take nominal values of the MC dimensions (for Sample I and II) and the saturation magnetization of YIG as $M_{S} = 0.143 \times 10^6$~A/m and $M_{S} = 0.142 \times 10^6$~A/m for Sample I and II, respectively. Note the difference in $M_{S}$ can be attributed to growth anisotropy fields which results from YIG and GGG lattice mismutch during epitaxy. The conductivity of the metal is assumed as $\sigma = 5.8 \times 10^7$ S/m, which is a tabular value for Cu.

\begin{center}
\begin{figure}[!ht]
\includegraphics[width=4cm]{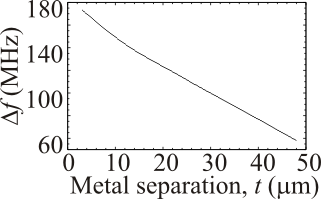}
\caption{The metallization induced frequency shift ($\Delta f$) of the magnonic band gap in Sample I in dependence on the separation distance of the metal from the magnonic crystal surface.}
\label{Fig:separation}
\end{figure}
\end{center}

Because the MC was putted just on the top of the transducers without any polishing of the MC surface, the real separation can be in the range of few to dozen $\mu$m due to some irregularities or dust grains (schematically shown in Fig.~\ref{structure}). In addition, any irregularity of YIG film or a metal layer will also results in an increase of this separation. It is also possible that conductivity of the metal $\sigma$ or its thickness $D$ are different than assumed in the calculations. However, the dispersion relation  is not sensitive for variation of the product of conductivity and thickness of the metal,~$\sigma D$ at a range of SSW wavectors in the interest. \cite{Filimonov2002} Thus only the $t$ was used as a free parameter in calculations. In Fig.~\ref{Fig:separation} the frequency shift ($\Delta f$)  of the first magnonic band gap in the Sample I with the metallic overlayer with respect to the MC without the metal is shown in dependence on $t$. There is a monotonous decrease of $\Delta f$ with increasing $t$, because the influence of the metal decreases.  The agreement with the experimental data for Sample I with the metallic overlayer is achieved for $t = 18$~$\mu$m and this value is used in the following calculations. 

In Fig.~\ref{results1} (e) the results of calculated dispersion relation are presented by blue and red dots for Sample I without and with metallic overlayer, respectively. In this figure we can see that the edge of the magnonic band gap in the metallized structure is shifted from the border of the Brillouin zone and this effect is due to nonreciprocity of the SSW dispersion relation. The gap opens when the interacting SSWs with equal frequencies fulfill the general Bragg condition.\cite{Mruczkiewicz2013b}
Magnonic band gaps are marked by transparent blue and red horizontal bars for measurements with and without metallic overlayer, respectively. The size of the first magnonic gap is $\Delta F \approx 39$ MHz for sample with metal overlayer and $\Delta F \approx 35$ MHz for sample without metal overlayer. The position and the width of the band gaps from calculations agree very well with the gaps found in the transmission experiment. Thus, we have shown experimentally that magnonic band gaps can exist in MCs covered by a metallic overlayer in the frequency and wavevector part of the SSW spectra where nonreciprocal properties are significant.

From the measured PF characteristic we can obtain even more information about the SSW dispersion in MCs than from the AF characteristic. The PF graphs are reduced to the $-\pi$ and $\pi$ values and of course the waves with phase difference $2\pi$ are not differentiable. But the phase jump over $2\pi$ phase difference visible in the PF characteristic at frequencies from a transmission band is determined by the separation distance between the transducers and is related to the change of the SSW wavevector $\Delta k= 2 \pi /L_{\text{T}} = 0.157 \times 10^4$ m$^{-1}$. This way the dispersion relation can be estimated, although  ambiguously it gives a clear information about the slope of the dispersion curve (i.e., its group velocity) at different parts of the magnonic spectra. Comparing Fig.~\ref{results1} (b) and (d) we can see that the group velocity of SSWs from the first and second band in MC with metal are much higher then respective velocities in MC without a metal. By assigning a wavevenumber $k=n \pi / a$ for the frequency corresponding to the value of phase equal to $- \pi$ at the edge of the $n$-th magnonic gap we can superimpose the function $f(k)$ extracted from PF characteristic  on the calculated dispersion relation. These dispersions are presented with red and blue continuous lines in Fig.~\ref{results1}(e) for Sample I. The agreement between measured and calculated dispersion curves is very good and proves that the metallic overlayer can be used not only to tune position of the magnonic band gap but also to mold the velocity of the transmitted signal. Moreover, the shape of the dispersion curve at the magnonic band gap edge can be important factor for possible applications of MCs, e.g., as ultrasensitive sensors of a magnetic field.\cite{Inoue11:132511}

Fig.~\ref{results2} (a)-(d) shows a results of measured AF and PF characteristics of unmetallized and metallized Sample II ($a=200$ $\mu$m  and width of the grooves $w=50$ $\mu$m). The comparison of dispersion relations extracted from  the PF characteristics with the results of calculations is shown in Fig.~\ref{results2} (e). In this case the separation between YIG and metal was taken as $t=28$ $\mu$m.  The AF and PF characteristics show wide band gaps (wider then for Sample I) at different frequency ranges. The size of the first magnonic gap is $\Delta F \approx 61$ MHz for sample with metal overlayer and $\Delta F \approx 44$ MHz for sample without metal overlayer. It is due to the larger ratio of groove depth to YIG thickness. Again, band gap at elevated frequencies is measured with the use of coplanar lines. The numerical dispersion relation confirm existence of a large indirect magnonic bang gap and a larger group velocity of SSWs in the Sample II with agreement with the dispersion extracted from the PF characteristic.

\begin{center}
\begin{figure}[!ht]
\includegraphics[width=6cm]{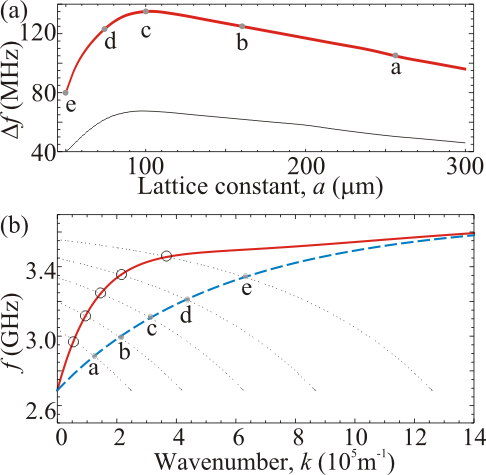}
\caption{(Color online) (a) Shift of the magnonic band gap frequency due to metallization as a function of the lattice constant. The thick red solid line is for YIG MC of 7.7 $\mu$m thickness and 1.5 $\mu$m deep grooves with the width of $w=0.5 a$. The black thin line is for YIG MC of thickness 3.8 $\mu$m and 2 $\mu$m deep grooves with the width of $w=0.25 a$. The separation of metal layer from MC was fixed to 18 $\mu$m in both cases.
(b) The SSW dispersion relation of a homogeneous YIG fim of 7.7 $\mu$m thickness with a metal overlayer (red solid line) and without metal (blue dashed line). The gray dotted lines are artificially introduced back-folded branches due to selected periodicities marked in (a) by letters from a to e. The small full-gray and big empty dots indicate the  expected positions of the magnonic band gap in MC without (which opens at the border of the first BZ, $\pi/a$) and with metallic overlayer (inside the BZ), respectively, for selected lattice constant in (a).  }
\label{results4}
\end{figure}
\end{center}

Additional calculations were performed to find influence of the lattice constant on the shift of the frequency of the magnonic band gap due to metallization. The frequency of the band gap shift in dependance on the lattice constant for YIG MC of 7.7 $\mu$m thickness and 1.5 $\mu$m deep grooves with the width of $w=0.5 a$ is shown with bold solid red line in Fig. \ref{results4} (a).   The same function is also plotted for MC of  thickness 3.8 $\mu$m and 2 $\mu$m deep grooves of the width $w=0.25 a$ with thin gray line. In both cases the separation of metal from MC surface was fixed to 18 $\mu$m. It can be seen that for MCs characterized by different thicknesses it is possible to define an optimal value of the lattice constant for which the shift of the band gap is largest.  

The dependance in Fig.~\ref{results4} (a) can be understood by  analysis of the dispersion relation of SSWs in a homogeneous YIG film shown in Fig.~\ref{results4} (b), where the solid-red and blue-dashed lines point at the film with and without metallic overlayer, respectively. The position of the band gap in MC can be estimated by introducing artificial periodicity into homogeneous film and the exploitation of the band periodicity in the wavevector space according with the Bloch theorem. Therefore, the thin-dotted lines in Fig.~\ref{results4} (b) show dispersion relations of SSW propagating in the $-k$ direction (i.e., unaffected by the metal film) shifted by the reciprocal lattice vector $2\pi/a$ for five selected lattice constants marked by letters a to e in Fig.~\ref{results4} (a). The small full-dots indicate the first BZ boundary (and the location of magnonic band gap in MC without metal) for chosen lattice constants; the big empty-dots point at the intersections of dispersion for $+k$ (solid red) with $-k+2\pi/a$ (i.e., the expected location of the magnonic band gaps) for the metallized MC. The observed increase of the band gap shift with decreasing lattice constant from $a=300$ $\mu$m to 100 $\mu$m (points a, b, c in Fig.~\ref{results4} (a))  is because of increasing nonreciprocity (i.e., an increase of the difference between the red-solid and dashed-blue line in Fig.~\ref{results4} (b)).
However, for  lattice constants smaller than $a=100$ $\mu$m a decrease of the band gap shift is observed although the nonreciprocity is still meaningful close to 100 $\mu$m. This is due to two effects. The first is the decrease of the nonreciprocity of the dispersion relation due to the finite conductivity and finite separation of the metal ovelayer. The second is a decrease of the group velocity of the magnonic branch that is not affected by the metal.  For MCs with lattice constant smaller than $a=73$ $\mu$m (point d in Fig.~\ref{results4}) the shift is decreasing rapidly, since the nonreciprocity effect disappears at large wavevectors. Thus, we can conclude that the maximal shift of the frequency band gap due to contact with the metallic film is at the value of lattice constant $a$ for which the first Brillouin zone boundary ($\pi / a$) is close to the wavenumber value where in homogeneous YIG film the nonreciprocity is largest.

 In summary, we have studied transmission of the surface spin waves in one-dimensional yttrium iron garnet films based magnonic crystal placed near the metal screen under conditions when incident and reflected waves have nonreciprocal dispersion  $f(k) \neq f(-k)$. We have proved experimentally previous theoretical findings that  nonreciprocity of surface spin waves do not prevent formation of the gaps in the magnonic spectra of the metallized magnonic crystal. Due to nonreciprocity the shift of the magnonic bandgaps to higher frequency and from the border of the Brillouin zone were detected.  The numerical calculations based on COMSOL software give very good agreement with experimental measurements of the magnonic bandgap shift $\Delta f$, bandgap width $\Delta F$ and dispersion characteristics. The value of the frequency shift $\Delta f$ of the magnonic gap depends on the magnonic crystal lattice constant, magnetic film and dielectric spacer thicknesses. For magnonic crystals based on typical yttrium iron garnet films shift of magnonic gap can be $\Delta f > 100$ MHz and up to several times greater than width of magnonic gap. From the last circumstances one can conclude that nonreciprocal waves are promising for manipulation of the spin wave band structure in magnonic crystals and for microwave signal processing. 
 

\begin{acknowledgments}
We acknowledge the financial assistance from the European Community, Grant No. FP7/2007-2013 under GA 247556 (People) NoWaPhen, from the National Science Centre of Poland, Project DEC-2-12/07/E/ST3/00538, from the Russian Foundation for Basic Research (13-07-00941-a, 14-07-00896-аa and 13-07-12421-ofi-m) and from the Grant of the Government of the Russian Federation (Contract No. 11.G34.31.0030). \end{acknowledgments}

\bibliographystyle{apsrev4-1}
\bibliography{bibliography3}

\end{document}